\documentclass[preprint,aps]{revtex4}

\usepackage{epsf}
\usepackage{color}
\usepackage{graphicx}

\begin{document}

\title{Ground-State Ordering of the $J_1$--$J_2$ Model
on the Simple Cubic and Body-Centered Cubic Lattices}

\author{D. J. J. Farnell}

\affiliation{School of Dentistry, Cardiff University, 
Cardiff CF14 4XY, Wales, United Kingdom}
\author{O. G\"otze and J. Richter 
}

\affiliation{Institut f\"ur Theoretische Physik, Universit\"at Magdeburg, 
D-39016 Magdeburg, Germany
}

\date{\today}

\begin{abstract}
The $J_1$--$J_2$ Heisenberg model is a ``canonical'' model in the field of quantum
magnetism in order to study the interplay between frustration and quantum fluctuations 
as well as quantum phase transitions 
driven by frustration. Here we apply the Coupled Cluster Method (CCM) to 
study the spin-half $J_1$--$J_2$ model with antiferromagnetic
nearest-neighbor bonds $J_1 >0$ and next-nearest-neighbor bonds $J_2 >0$
for the simple cubic (SC) and 
body-centered cubic (BCC) lattices. In particular, we wish to study the ground-state 
ordering of these systems as a function of the frustration parameter 
$p=z_2J_2/z_1J_1$, where $z_1$ ($z_2$) is the number of nearest
(next-nearest) neighbors. We wish to determine the positions of the 
phase transitions using the CCM and we aim to 
resolve the nature of the phase transition points.
We consider the ground-state energy, order parameters, spin-spin
correlation functions as well as the spin stiffness in order to determine the ground-state phase diagrams
of these models. We find a direct first-order phase transition at a value of 
$p = 0.528$ from a state of nearest-neighbor 
N\'eel order to next-nearest-neighbor N\'eel order for the BCC  lattice. 
For the SC lattice the situation is more subtle.  
CCM results for the energy, the order parameter, 
the spin-spin correlation functions and the spin stiffness indicate that
there is no direct first-order transition between ground-state phases
with magnetic long-range order, rather it is more likely that 
two phases
with antiferromagnetic long-range are separated by a narrow region of
a spin-liquid like quantum phase around $p=0.55$.
Thus the strong frustration present  in the
 $J_1$--$J_2$ Heisenberg model on the SC lattice may open a window for an
 unconventional quantum ground state in this three-dimensional spin model.
\end{abstract}

\maketitle

\section{Introduction}

Frustrated quantum magnetism continues to attract enormous attention both in
theory and experiment\cite{lnp645,lacroix,balents}.
A canonical model to study the interplay of frustration and quantum fluctuations
is the spin-half $J_1$--$J_2$ Heisenberg model. On the square lattice this
model 
has been extensively utilized to study frustration-driven quantum 
phase transitions between semiclassical ground-state phases with magnetic 
long-range order and magnetically disordered quantum phases, see, e.g., 
Refs.
\cite{henley,ref01,ref02,ref03,ref04,ref05,ref06,ref07,ref08,ref09,ref010,ref011,ref012,ref013,ref014,ref015,ref016,ref017,ref018,ref019,ref020,ref021,ref022,ref023,ref024,ref025,ref026}. 
Despite of the numerous investigations of the two-dimensional (2D) model the nature of the
non-magnetic quantum
phase around $J_2/J_1  =0.5$ is still under debate.
Interest in the spin-half $J_1$--$J_2$ model on square lattice is 
motivated also by its relation to experimental studies of various magnetic 
materials, such as VOMoO$_4$ (Ref. \cite{ref15}), Li$_2$VOSiO$_4$, 
and Li$_2$VOGeO$_4$ (Ref. \cite{ref16}) or Sr$2$CuTeO$_6$ (Ref.
\cite{tanaka2014}).

\begin{figure}
\epsfxsize=9cm
\centerline{\epsffile{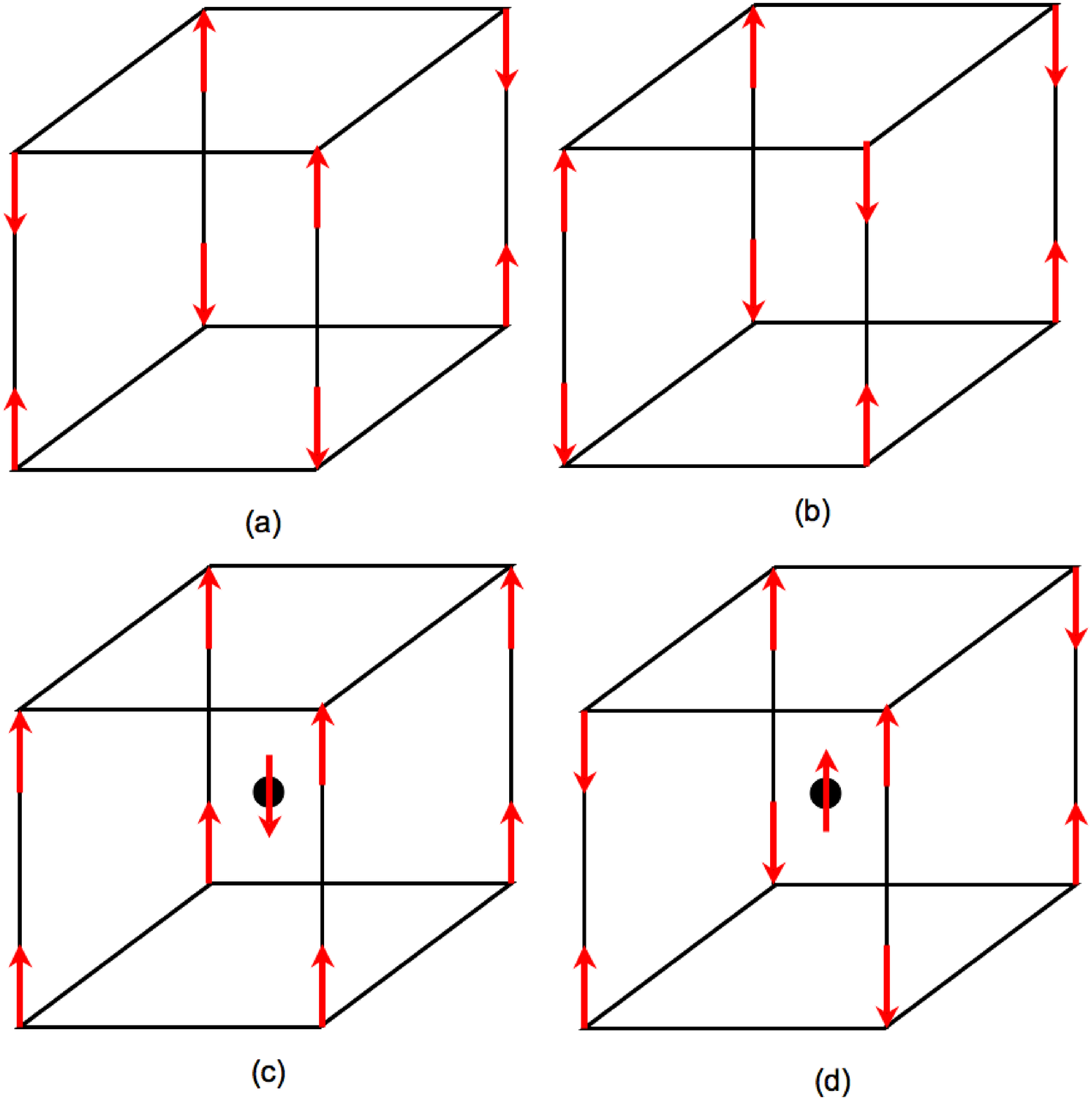}}
\caption{CCM model states: (a) N\'eel model state for the simple cubic lattice
(denoted by SC-AF1); (b) 
striped model state for the  simple cubic lattice (denoted by SC-AF2); (c) nearest-neighbor N\'eel model 
state for the body-centered cubic lattice (denoted by BCC-AF1); (d) next-nearest-neighbor N\'eel
striped model 
state for the body-centered cubic  lattice (denoted by BCC-AF2).}
\label{fig1}
\end{figure}

The dimension of  the underlying lattice is crucial to the existence of magnetic 
long-range order in quantum magnetic systems. 
Naturally there is a stronger tendency to order in three-dimensional (3D)
systems.
Thus, already a quite small coupling between the $J_1-J_2$ square-lattice layers
leads to a disappearance of the magnetically disordered phase
\cite{Schm:2006,rojas2011,Holt2011,Fan2014}.
However,  a magnetically disordered
quantum phase is not per se excluded in frustrated 3D systems, as it has been
demonstrated
for the spin-half Heisenberg antiferromagnet (HAFM) on
the pyrochlore lattice \cite{canals98}.

The natural 3D counterpart of the square-lattice $J_1$-$J_2$ model is the $J_1$-$J_2$ model
on the body-centered cubic (BCC) lattice.  The limiting case of $J_1=0$ and
$J_2>0$ belongs to the case of two interpenetrating 
unfrustrated, i.e. bipartite, antiferromagnets for both models.
The few investigations of the 3D BCC spin-half  $J_1$-$J_2$ model include 
exact diagonalization (ED) \cite{ref25}, series expansions around the Ising
limit \cite{ref23}, spin-wave theory \cite{ref25,ref24},  and 
the random phase approximation \cite{ref26}. Thus, all methods (except ED) 
start from the symmetry-broken classical antiferromagnetic  states and then quantum corrections
are subsequently taken into account. 
Consistently,  all of these methods indicate that   
a single phase transition occurs in this system. In contrast to the 2D model, a
magnetically disordered quantum phase is not observed. 
However, the frustration  has a strong influence on the thermodynamics, in
particular the
critical temperature is substantially suppressed by frustration
\cite{pinettes,ref23,bcc_TD_EBJB2015,bcc_ICM2015}.

Less clear is the situation for the spin-half $J_1$-$J_2$ model
on the simple cubic (SC) lattice
\cite{pinettes,ref16a,ref17,barabanov1995,ref20,ref21,ref22,thomale2015}.
In this case different approaches, such as, 
spin-wave theories \cite{ref16a,ref17,ref20,ref22}, variational 
cluster approach \cite{thomale2015}, differential  
operator technique \cite{ref21} or a spherically symmetric Green function
method \cite{barabanov1995}, 
come to different conclusions with respect
to the existence of a disordered ground-state phase.
The underlying semi-classical physics of these approaches is
different. Spin-wave theories \cite{ref16a,ref17,ref20,ref22}, differential  
operator technique \cite{ref21}, and the variational
cluster approach \cite{thomale2015} include explicit symmetry 
breaking. Spin-wave theory uses the $z$-axis aligned classical 
states as a starting point for the calculation, whereas differential
operator technique and the variational cluster approach
use Weiss fields to test the presence of the antiferromagnetic order. 
By contrast, the Green function method \cite{barabanov1995} 
preserves full spin rotational invariance. 
A direct first-order transition between two antiferromagnetically long-range
ordered phases was obtained in
Refs.~\cite{ref16a,ref17,ref21,ref22}, whereas within Green
function technique \cite{barabanov1995} and  linear spin-wave theory
\cite{ref22} a magnetically  disordered quantum
phase was found that separates the  two antiferromagnetic phases.         
Very recently the role of a third-neighbor coupling, $J_3$, was studied
by Laubach et al. \cite{thomale2015}. Although, these authors did not
discuss
a disordered  quantum phase for $J_3=0$, their results indicate that a very small 
additional frustrating   $J_3>0$ leads to such a spin-liquid like
quantum phase. 
It is in order to emphasize the basic 
difference between the BCC and SC $J_1$-$J_2$ models, that becomes evident in the limit of large
$J_2$ (or $J_1 \to 0$). Contrary to the BCC model, the $J_1$-$J_2$ HAFM on
the SC model is still strongly frustrated, because the antiferromagnetic $J_2$ bonds connect
sites of two interpenetrating face-centered cubic (FCC) lattices.

In the present paper we use  the coupled cluster method (CCM) to perform a
comparative 
study of the spin-half  $J_1$-$J_2$ HAFM
on the BCC and SC lattices.
We mention here, that the CCM previously 
has been applied to the 2D  square-lattice $J_1$-$J_2$ HAFM 
\cite{ref06,ref010,ref012,ref013,ref022,J1J2_FM_CCM} and the method provides accurate
results for
the  ground-state energy, the magnetic order parameter as well as
for the critical points, where the quantum phase transitions take place.

The relevant Hamiltonian of the $J_1$-$J_2$ model is given by
\begin{equation}
H = J_1 \sum_{\langle i,j \rangle } {\bf s}_i \cdot {\bf s}_j +
J_2 \sum_{\langle \langle i,j \rangle \rangle } {\bf s}_i \cdot {\bf s}_j ~~ .
\label{eq1}
\end{equation}
The symbol $\langle i,j \rangle$  indicates those bonds that connect nearest-neighbor 
sites (counting each bond once only) and the symbol $\langle \langle i,j \rangle \rangle$  indicates 
those bonds that connect next-nearest-neighbor sites (again counting each bond 
once only). Here we consider the SC and BCC lattices in the regime $J_1 \ge 0$ and 
$J_2 \ge 0$, and these lattices (and CCM ``model states'', see
Sec.~\ref{method}) are shown in Fig. \ref{fig1}.
We note that these systems are frustrated by positive values of $J_2$. The
competition between the bonds $J_1$ and $J_2$ and therefore the phase 
transition points in these systems depend on coordination numbers $z_1$ (i.e., the number 
of nearest-neighbors) and $z_2$ (i.e., the number of next-nearest-neighbors). In 
order to enable our calculations to be consistent with each other, we introduce the 
following quantity, 
\begin{equation}
p = \frac {J_2 z_2} {J_1 z_1} ~~ .
\label{eqp}
\end{equation}
The (underlying) BCC and SC lattices are both bipartite, and so the 
nearest-neighbor N\'eel state forms the classical ground state for both
of these systems for smaller values of $p<p_{\rm cl}$, i.e., up to 
the phase transition point at $p=p_{\rm cl}$, where $p_{\rm cl}=\frac 12$
for the SC as well as for the  BCC  lattice.
 These states are shown 
in Fig. \ref{fig1} for both the SC and BCC  lattices. They are denoted by
 SC-AF1 and BCC-AF1, respectively.
The situation is 
more complicated in the large $p$ limit. The BCC lattice decouples 
into two SC lattices when nearest-neighbor bonds are set to $J_1=0$
and $J_2$ remains non-zero. Thus, collinear striped order (the corresponding
state is denoted by BCC-AF2)
 occurs for 
$p>p_{\rm cl}$ for the BCC lattice, also shown in Fig. \ref{fig1} for the BCC 
lattice. We shall use this state as another model state for the BCC lattice. 
By contrast, the SC lattice decouples 
into two FCC lattices when nearest-neighbor bonds are set to $J_1=0$
and $J_2$ remains non-zero. This system (with only next-nearest-neighbor 
antiferromagnetic bonds) is therefore frustrated and there is  a highly
degenerate  classical ground-state manifold including non-collinear ground
states. However, according to the {\it order by disorder} mechanism \cite{villain,
shender} 
collinear striped ordering is favored by quantum fluctuations
\cite{ref16a,ref17,barabanov1995,ref20,ref21,ref22,thomale2015} 
also for $p > p_{cl}$. The ``striped'' model state for the SC lattice (denoted
by SC-AF2)
used here is also shown  in Fig. \ref{fig1}.

Here we wish to investigate the ground-state properties of the 
spin-half $J_1$--$J_2$ model on the SC and BCC 
lattices by using the CCM. We wish to determine the positions of the 
phase transitions using the CCM and we aim to 
discuss the nature of the phase transitions. As there is arguably 
less evidence available in the literature for the SC lattice rather than 
the BCC lattice, this investigation should be most useful for the SC 
lattice. However, we shall see that insight into both systems can 
be obtained by comparing and contrasting the results for each 
system.

In what follows, the formalism of the CCM is presented briefly, 
and then the results for the BCC lattice and the SC 
lattice are given. We present our conclusions in the final section of 
this paper. 

\section{Method}
\label{method}
For  general information relating to the methodology of the CCM, see, e.g.,
Refs.~\cite{ccm_theory,zeng98,bishop98a,bishop04,spin_systems_book}. The 
CCM has recently been applied computationally at high orders of approximation 
to quantum magnetic systems with much success, see, e.g., 
Refs.~\cite{spin_half_xxz,ccm_j_prime,ccm_shastry,ccm_extra,kagome_general_s,honeyj1j2j3,archi2014,kagome_XXZ,Kagome_layer,jiang2015a,jiang2015b,bishop2015}.  In the field of quantum magnetism, 
advantages of this approach are that it can be applied 
to strongly frustrated quantum spin systems in any dimension and 
with arbitrary spin quantum numbers. 
The exact ket and bra ground-state energy 
eigenvectors, $|\Psi\rangle$ and $\langle\tilde{\Psi}|$, of a 
many-body system described by a Hamiltonian $H$, 
\begin{equation} 
H |\Psi\rangle = E_g |\Psi\rangle
\;; 
\;\;\;  
\langle\tilde{\Psi}| H = E_g \langle\tilde{\Psi}| 
\;, 
\label{eqH} 
\end{equation} 
are parametrized within the CCM as follows:   
\begin{eqnarray} 
|\Psi\rangle = {\rm e}^S |\Phi\rangle \; &;&  
\;\;\; S=\sum_{I \neq 0} {\cal S}_I C_I^{+}  \nonumber \; , \\ 
\langle\tilde{\Psi}| = \langle\Phi| \tilde{S} {\rm e}^{-S} \; &;& 
\;\;\; \tilde{S} =1 + \sum_{I \neq 0} \tilde{{\cal S}}_I C_I^{-} \; .  
\label{eq2} 
\end{eqnarray} 
Again, we remark that the model or reference states $|\Phi\rangle$ 
for the SC and BCC lattices are shown in Fig. \ref{fig1}. The ground-state 
energy is now given by
\begin{equation} 
E_g = E_g ( \{{\cal S}_I\} ) = \langle\Phi| {\rm e}^{-S} H {\rm e}^S|\Phi\rangle
\;\; . 
\label{eq9}
\end{equation}  
The ket-state and bra-state correlation coefficients are obtained by solving the CCM ket- and bra-state equations given by
\begin{eqnarray}
\label{ket_eq}
\langle\Phi|C_I^-e^{-S}He^S|\Phi\rangle = 0,
\qquad 
\forall I\neq 0,
\\
\label{bra_eq}
\langle\Phi|\tilde{\cal S}e^{-S}[H, C_I^+]e^S|\Phi\rangle = 0,
\qquad 
\forall I\neq 0.
\end{eqnarray}
Each ket- or bra-state equation belongs to a certain creation  operator $C_I^+=s_i^+,\,\,s_i^+s_{j}^+,\,\, s_i^+s_{j}^+s_{k}^+,\cdots$, i.e. it corresponds to a certain set (configuration or cluster) of lattice sites $i,j,k,\dots\;$. The ket- and bra-state correlation coefficients ${\cal S}_I$ and  $\tilde{\cal S}_I$, respectively, relate to the ``fundamental'' cluster with index $I$ (of $N_f$ such fundamental clusters in total) and so also to the appropriate ground-state equation above.

The manner in which is the CCM equations are determined and solved is 
discussed elsewhere (again, see, e.g.,
Refs.~\cite{spin_half_xxz,ccm_j_prime,ccm_shastry,ccm_extra,kagome_general_s,honeyj1j2j3,archi2014,kagome_XXZ,Kagome_layer,jiang2015a,jiang2015b,bishop2015} for more details).
However, it is important to note here that the CCM formalism 
is only ever exact in the limit of inclusion of
all possible multi-spin cluster correlations within 
$S$ and $\tilde S$, although in any real application 
this is usually impossible to achieve. It is therefore 
necessary to utilize various approximation schemes 
within $S$ and $\tilde{S}$. The most commonly 
employed scheme has been 
the localized LSUB$m$ scheme, 
in which all multi-spin correlations over distinct 
locales on the lattice defined by $m$ or fewer contiguous 
sites are retained. We will use this scheme in this article.

Note that we also 
make the specific and explicit restriction that the 
creation operators $\{C_I^+\}$ in $S$ preserve the 
relationship that, in the original (unrotated) spin
coordinates,  $s^z_T=\sum_i s^z_i=0$ in order to 
keep the approximate CCM ground-state wave function in
the correct ($s^z_T=0$) subspace. Note that each 
fundamental cluster is independent of all others clusters 
with respect to the symmetries of the lattice (and Hamiltonian). 

The order parameter (sublattice magnetization) $M$ for the 
systems considered here is defined as
\begin{equation}
M = -\frac 1{N} \sum_i^N \langle \tilde \Psi | {\hat s}_i^z | \Psi \rangle ~~ ,
\label{eq11}
\end{equation}
where we note that ${\hat s}_i^z$ is with respect to the local spin axes 
at site $i$ {\it after rotation} of the local spin axes with respect to 
the model state so that ({\it notationally only}) the spins appear to 
align in the negative $z$-direction. This ensures that the mathematics
of treating these problems is slightly simpler 
\cite{spin_systems_book,spin_half_xxz}. Hence, the order
parameters are taken with respect to the model states shown
in Fig. \ref{fig1}. 

As mentioned above, the LSUB$m$ approximation becomes exact only in the limit $m 
\rightarrow \infty$, and so it is useful to extrapolate the LSUB$m$ 
results in this limit. A well-established extrapolation
scheme \cite{spin_systems_book,spin_half_xxz,ccm_j_prime,ccm_shastry,ccm_extra,kagome_general_s,honeyj1j2j3,archi2014,kagome_XXZ,Kagome_layer,jiang2015a,jiang2015b,bishop2015}
 for the 
ground-state energy, $E_g/N$, is given by
\begin{equation}
e_g(m)=E_g(m)/N = e_{g}(m = \infty) + a_{1}m^{-2}+a_{2}m^{-4}\, .     \label{E_extrapo}
\end{equation}
For the magnetic order parameter $M$ 
we use the scheme
\begin{equation}
M(m) = M(m=\infty) + b_1/m^{1/2}
+ b_2/m^{3/2} .  \label{M_extrapo_frustrated}
\end{equation}
This extrapolation ansatz is most suitable to detect 
ground-state order-disorder
transitions \cite{ref012,ref013,ref022,ccm_extra,kagome_general_s,honeyj1j2j3,archi2014}.
We were able to carry out CCM calculations to the LSUB8 level of approximation 
for the BCC lattice and to LSUB10 for the SC lattice.
The maximum number of fundamental configurations entering the CCM
calculations  at the  LSUB10  level of approximation is $1,728,469$.

We know from
Refs.~\cite{ref012,ref013,ref022,ccm_extra,kagome_general_s,honeyj1j2j3,archi2014} 
that the lowest level of approximation, LSUB2, conforms poorly to the extrapolation 
schemes, especially as the parameter $p$ increases. Hence, as in previous calculations,
we exclude LSUB2 data from the extrapolations.

Specifically for the SC lattice we will also calculate the spin stiffness $\rho$
up to the LSUB8 level of approximation. More explanation is needed relating 
to how to define the stiffness and how to perform the necessary CCM calculations, 
and so we transfer this discussion to the Appendix~\ref{app_rho}.


\begin{figure}
\epsfxsize=12cm
\centerline{\epsffile{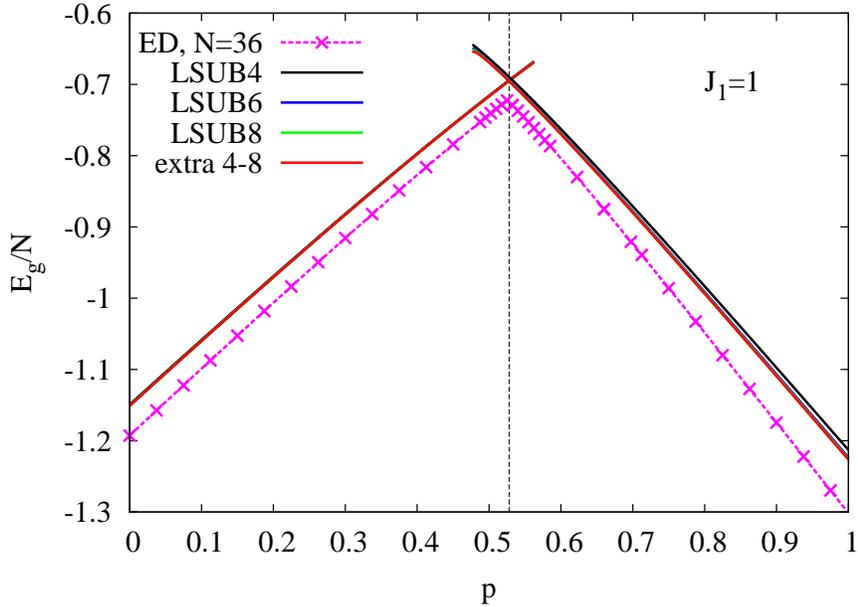}}
\caption{CCM results for the ground-state energy of the spin-half $J_1$--$J_2$ model  
on the BCC lattice are compared to results of exact diagonalizations (ED) with $N=36$.
Note that the curves for the LSUB$m$ data coincide almost completely.  
Extrapolated results (label `extra 4-8') are obtained by using the
extrapolation scheme
of Eq.~(\ref{E_extrapo}) using data from the LSUB4, LSUB6, and LSUB8
approximations. 
The ground-state energies of the two model states are found to intersect at $p_c=0.528$.}
\label{fig2}
\end{figure}

\begin{figure}
\epsfxsize=12cm
\centerline{\epsffile{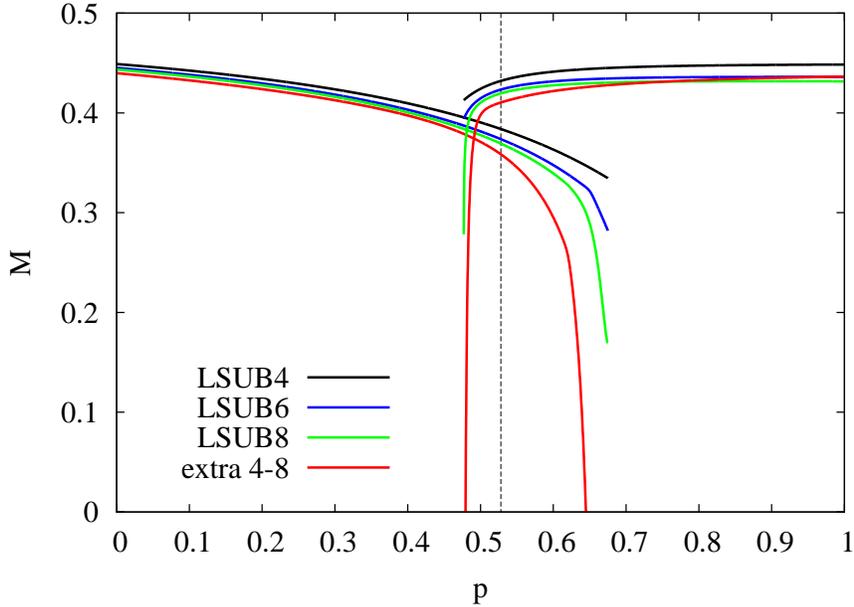}}
\caption{CCM results for the order
parameter (sublattice magnetization) $M$ of the spin-half $J_1$--$J_2$ model  
on the BCC lattice.
Extrapolated results (label `extra 4-8') are obtained by using the
extrapolation scheme
of Eq.~(\ref{M_extrapo_frustrated}) using data from the LSUB4, LSUB6, and LSUB8
approximations.
The vertical (dotted) line indicates the intersection point of the ground-state
energies for the two model states at $p_c=0.528$.}
\label{fig3}
\end{figure}

\begin{figure}
\epsfxsize=12cm
\centerline{\epsffile{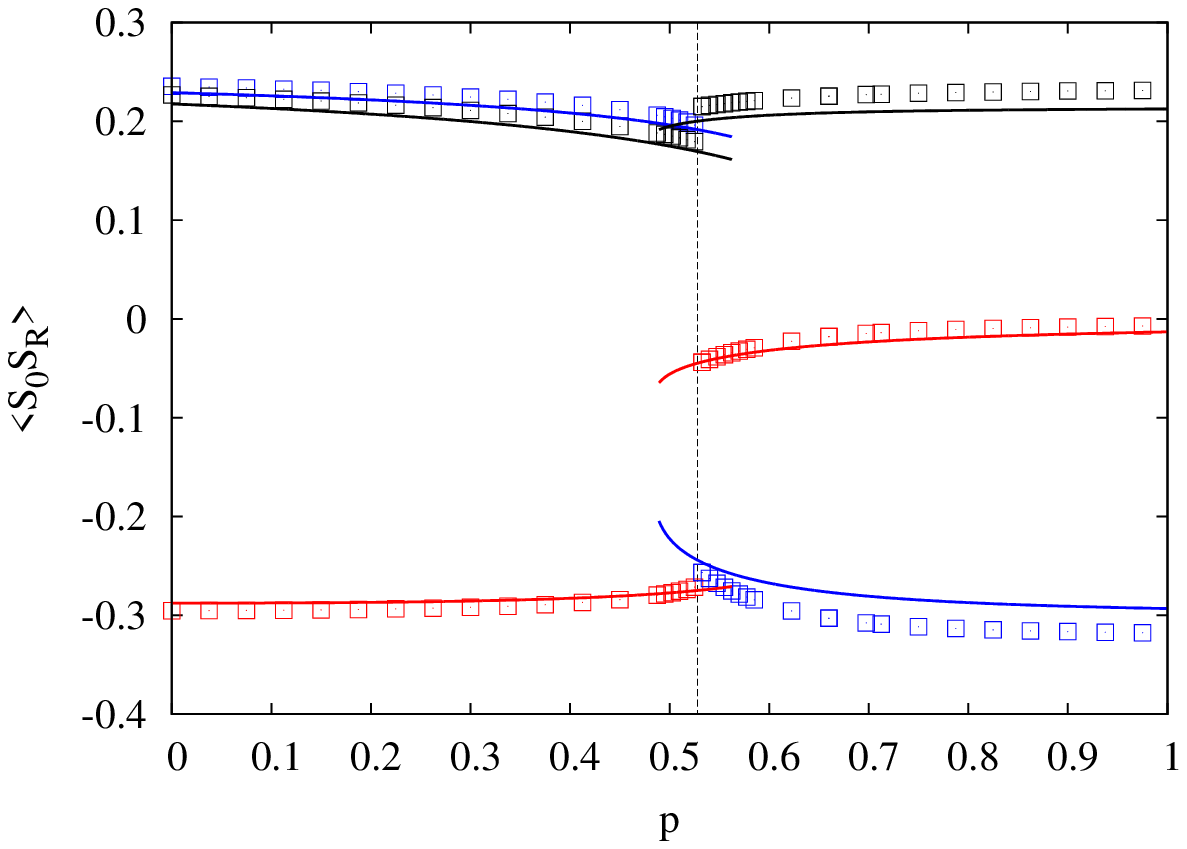}}
\caption{Spin-spin correlation $\langle {\bf S}_0{\bf S}_{\bf R}\rangle$
functions for nearest neighbors (red), next-nearest neighbors (blue) and for
third-nearest neighbors (black)  for the spin-half $J_1$--$J_2$ model on the
BCC lattice  in dependence of the frustration parameter $p=3J_2/4J_1$ (solid
lines - CCM-LSUB8 results, symbols - ED results for $N=36$, cf.
Ref.~\cite{ref25}).
All results are averaged data over all neighbors with the same separation
$|{\bf R}|$.   
}
\label{fig4}
\end{figure}

\section{Results}

\subsection{Body-Centered Cubic Lattice}

The BCC lattice is considered firstly. We were able to  
carry out CCM calculations to the LSUB8 level of approximation for this system.   
Results for the ground-state energy are shown in Fig. \ref{fig2}. 
LSUB$m$ results converge very rapidly with increasing level of approximation $m$, 
and differences in energies between LSUB6 and LSUB8 levels of approximation 
are broadly of order $10^{-4}$ for the BCC-AF1 model state and of order $10^{-3}$ 
for the BCC-AF2 model state and for all values of $p$. LSUB4, 
LSUB6, and LSUB8 results for the unfrustrated case where $p=0$ (setting also $J_1=1$) are given by 
$e_g = -1.14950$, $-1.15072$, and $-1.15101$, respectively. The
extrapolation to $m = \infty$ yields $e_g = -1.1513$, which  
compares well to 
results of series expansions of $e_g = -1.1510$  \cite{BCC_heisenberg_series_swt}
and of third-order spin-wave theory of $e_g = -1.1512$  \cite{BCC_heisenberg_series_swt}.
Good correspondence with ED results of Ref. \cite{ref25} are also seen by visual
inspection of Fig. \ref{fig2}. We observe that CCM and ED results follow a very similar 
pattern as we increase $p$, although ED results are clearly much lower in energy than 
those of the CCM. The difference between ED and CCM results is due to the finite 
size of the lattice ($N=36$) in the ED calculations. 
The overall behavior of the ground-state energy  provides clear evidence for
a first-order transition.  The intersection point at $p=p_c = 0.528$ of the ground-state energies of the BCC-AF1 and
BCC-AF2 energies determines the transition point.  
The corresponding kink in the $e_g(p)$-curve for $N=36$ (ED)
is at $p \approx 0.525$.

Results for the order parameter are shown in Fig.~\ref{fig3}. We see again that CCM 
results are converging with increasing level of LSUB$m$ approximation level, albeit 
more slowly than for the ground-state energy. LSUB4, LSUB6, and
LSUB8 results for
the unfrustrated HAFM (i.e., when 
$p=0$) are given by $M = 0.44899$, $0.44515$, and $0.44350$ respectively, and
the extrapolated value is 
$M=0.4398$.
Again, this result 
compares quite well to those predictions of series expansions of $M = 0.442$ 
\cite{BCC_heisenberg_series_swt}  and of third-order spin-wave theory of 
$M = 0.4412$ \cite{BCC_heisenberg_series_swt}. 
The data shown in  Fig.~\ref{fig3} clearly support that there is a direct
first-order transition between the phases with semi-classical magnetic
long-range orders of type AF1 and AF2 (see Fig.~\ref{fig1}).      
The values of the extrapolated order parameter at the transition point $p_c=0.528$ are
$M=0.3585$ (AF1) and
$M=0.4104$ (AF2).

The results for the spin-spin correlation functions at the LSUB8 level of approximation 
shown in Fig. \ref{fig4} agree well with the ED data for $N=36$. 
The change in the spin-spin correlation functions 
is very abrupt and the large magnitude of correlation functions
at $p = p_c$ is a further evidence of a first-order phase transition 
at this point. The small magnitude of the nearest-neighbor   spin-spin correlation
function at $p>p_c$ signals the splitting of the system in two weakly coupled
interpenetrating antiferromagnets with leading coupling $J_2$.

We may compare  the transition point $p_c=0.528$ obtained by the CCM
with previous results, namely $p_c=0.525$ (ED
\cite{ref25}), $p_c \approx  0.53$ (series expansions \cite{ref23} and 
non-linear spin-wave theory \cite{ref24}), $p_c  \approx  0.54$  (random phase approximation\cite{ref26}). Note that 
the critical point for the quantum model is slightly above the classical
value $p_{\rm cl} = 0.5$.

Finally, we emphasize the basic difference to the 2D square-lattice model
(see also the discussion in the next section).
Although, both models are of similar character concerning the competition
between the  $J_1$ and $J_2$ bonds, the increase  in dimension leads to a
significant stabilization of semi-classical magnetic long-range order   and
to the
disappearance of the intermediate quantum phase that is present in the 2D model.
Thus, the amount of frustration must be larger in 3D for such a magnetically 
disordered quantum phase to exist at all. 
The $J_1$-$J_2$ model on the SC lattice discussed in the next section 
might have a sufficient degree of frustration because the next-nearest-neighbor
bonds $J_2$ in this model compete not only with the nearest-neighbor bonds 
$J_2$ but also with each other.

\begin{figure}
\epsfxsize=14cm
\centerline{\epsffile{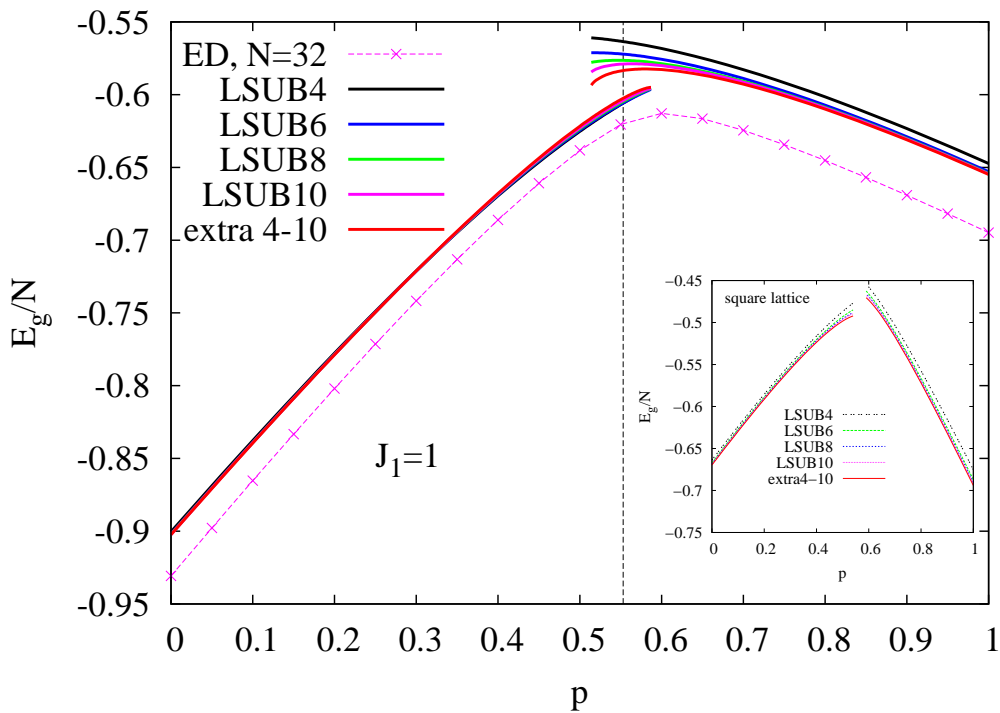}}
\caption{CCM results for the ground-state energy $E_g/N$ for the spin-half $J_1$--$J_2$ model  
on the SC lattice are compared to results of exact diagonalizations (ED) with $N=32$. 
Note that the curves for the LSUB$m$ data obtained for the N\'eel model state coincide almost completely.   
Extrapolated results (label `extra 4-10') are obtained by using the extrapolation scheme  
of Eq.~(\ref{E_extrapo}) using data from the LSUB4, LSUB6, LSUB8, and LSUB10 
approximations.
The vertical (dotted) line indicates the value in the middle of the two
points,  $p^{AF1}_c=0.549$ and $p^{AF2}_c=0.557$, where the extrapolated
order parameters of the SC-AF1 and SC-AF2 phases vanish. 
(Inset: CCM results for the spin-half square-lattice $J_1$--$J_2$ model
corresponding to Ref.~\cite{ref013}.)}
\label{fig5}
\end{figure}

\begin{figure}
\epsfxsize=14cm
\centerline{\epsffile{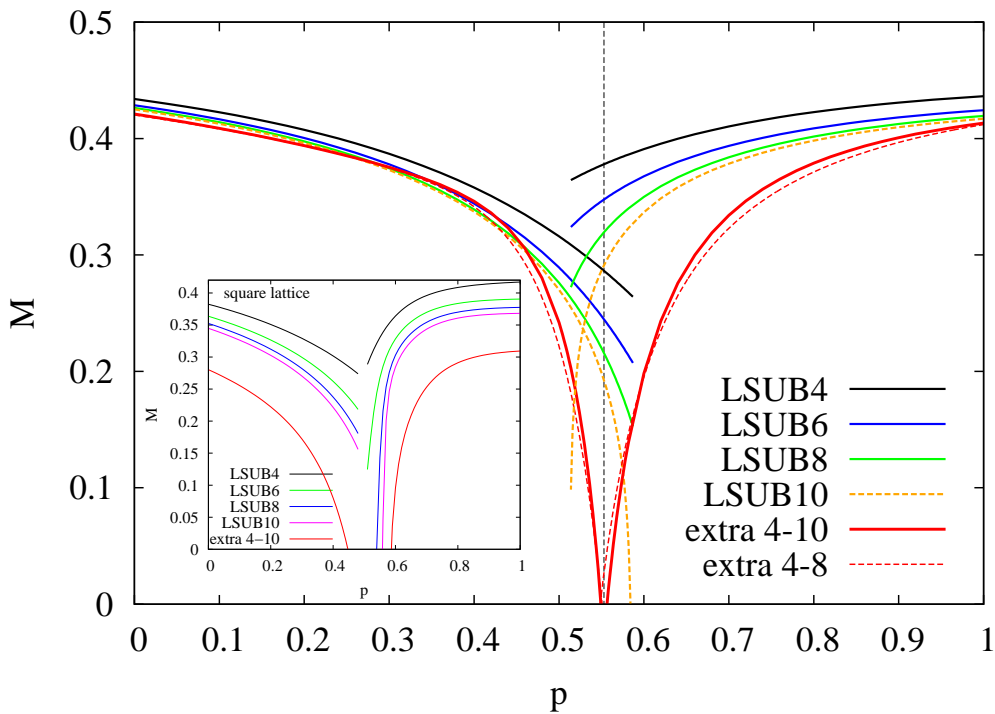}}
\caption{CCM results for the order parameter (sublattice magnetization) $M$
of the spin-half $J_1$--$J_2$ model  on the SC lattice.
Extrapolated results  are obtained by using the scheme of
Eq.~(\ref{M_extrapo_frustrated}). To get an impression of the accuracy of
the extrapolated order parameter we 
take into account (i) data from the LSUB4, LSUB6, LSUB8, and LSUB10
approximations (thick solid red line, label `extra 4-10') and (ii)        
 data from the LSUB4, LSUB6, and LSUB8 approximations (thin dotted red line,
 label `extra 4-8'). Obviously, both red lines are very close to each other.
The vertical (dotted) line indicates the value in the middle of the two
phase transition points  $p^{AF1}_c=0.549$ and $p^{AF2}_c=0.557$.
(Inset: CCM results for the spin-half square-lattice $J_1$--$J_2$ model 
corresponding to Ref.~\cite{ref013}.)}
\label{fig6}
\end{figure}

\begin{figure}
\epsfxsize=12cm
\centerline{\epsffile{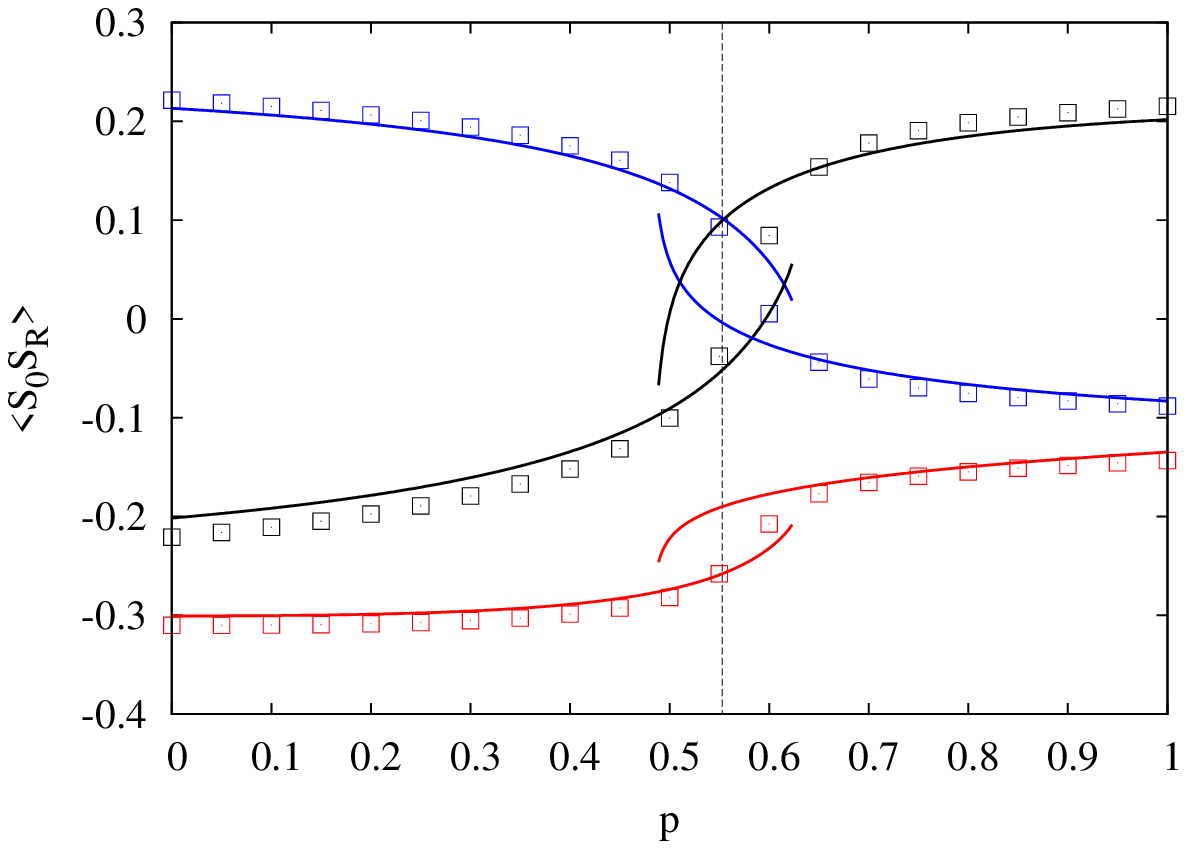}}
\caption{Spin-spin correlation $\langle {\bf S}_0{\bf S}_{\bf R}\rangle$
functions for nearest neighbors (red), next-nearest neighbors (blue) and for
third-nearest neighbors (black)  for the spin-half $J_1$--$J_2$ model on the
SC lattice  in dependence of the frustration parameter $p=3J_2/4J_1$ (solid
lines - CCM-LSUB8 results, symbols - ED results for $N=32$).
All results are averaged data over all neighbors with the same separation
$|{\bf R}|$.   
}
\label{fig7}
\end{figure}

\subsection{Simple Cubic Lattice}
Next we consider the SC lattice. We were able to  
carry out CCM calculations to the LSUB10 level of approximation for this system.    
Results for the ground-state energy on the SC lattice are shown in Fig.
\ref{fig5}. 
LSUB$m$ results are essentially converged at the LSUB10 level of approximation 
for the N\'eel model state SC-AF1 (differences in energy between the LSUB8 and
LSUB10
levels of approximation are generally much less than $10^{-3}$ for all values of
$p$.) 
Results for the striped model state SC-AF2 (only) do not demonstrate quite the same level 
of convergence as those results for the SC-AF1 N\'eel model state, although they are 
still close to each other.
For the unfrustrated  SC HAFM (i.e., when $p=0$ and  setting also $J_1=1$) LSUB4, LSUB6, LSUB8, and  LSUB10 results 
are $e_g =  -0.90043$, $-0.90180$, $-0.90214$,  and $-0.90225$, respectively.
We find an extrapolated CCM result of $e_g = -0.9024$
which compares well to 
results of series expansions of $e_g = -0.9021$  \cite{BCC_heisenberg_series_swt}
and  of third-order spin-wave theory of $e_g = -0.9025$ \cite{BCC_heisenberg_series_swt}.
Good correspondence with ED results  is again seen by visual inspection of Fig.
\ref{fig5},
although the difference between ED and CCM results is again due to the finite 
size of the lattice ($N=32$) in the ED calculations.

The curvature of the $e_g(p)$ curve around $p=0.55$ is noticeably different to the 
results for the ground-state energy for the BCC lattice near to its transition point. 
Moreover, we find that the solution to the LSUB10 equations on the SC lattice terminates at $p \sim 0.58$ 
for the SC-AF1 model state tracing and at $p \sim 0.52$ for the SC-AF2 
model state (i.e., we cannot trace the CCM solution beyond these termination points). 
One may expect that any intersection should occur within the region $0.52 \lesssim p \lesssim 0.58$, see Fig.5. 
However, a (tentative) extension of the ground-state energy for SC-AF1 model state 
beyond $p \sim 0.58$ with respect to $p$ until it crosses those results for the 
SC-AF1 model state leads to a speculative crossing point at $p \approx 0.65$, 
which is therefore clearly too large. We mention again that the energies for 
the N\'eel and striped model states demonstrate a very clearly defined 
intersection at $p_c$ for the BCC case, see Fig.2. 
On the other hand, the behavior of the ground-state energy of the spin-half square-lattice $J_1$--$J_2$ model, as is shown by the inset to Fig. \ref{fig5}, is quite similar to that of the SC lattice.

Results for the order parameter $M$ are shown in Fig.~\ref{fig6}. We see again that CCM 
results converge with increasing level of LSUB$m$ approximation level.
In order to provide an idea of the precision of the extrapolation of the order
parameter according to Eq.~(\ref{M_extrapo_frustrated}) two
extrapolation schemes are presented in  Fig.~\ref{fig6} : 
(i) data from the LSUB4, LSUB6, LSUB8, and LSUB10
approximations are used for the extrapolation
and 
(ii)    
only data from the LSUB4, LSUB6, and LSUB8 approximations are used. 
The results obtained by scheme (i) should be regarded
as more accurate than scheme (ii) because it contains 
more data to extrapolate with and higher orders of approximation. 
However, the differences in extrapolated results between both schemes 
remain small in the entire parameter region.
LSUB2, LSUB4, LSUB6, LSUB8, and LSUB10 results for the unfrustrated HAFM
(i.e. when $p=0$) 
are $M = 0.45024$,  $0.43392$,  $0.42860$,  $0.42626$, and 
$0.42504$, respectively. We find an extrapolated CCM result of
$M=0.4210$ ($M=0.4164$) for scheme (i) [scheme (ii)], and this result compares reasonably well to 
results of series expansions of $M= 0.424$ \cite{BCC_heisenberg_series_swt}
and of third-order spin-wave theory of $M = 0.4227$ \cite{BCC_heisenberg_series_swt}.

A striking difference to the BCC case is shown by the critical points 
that are estimated by finding the values at which the 
extrapolated order parameter becomes zero. We find $p_c^{AF1}=0.549$  
and $p_c^{AF2}=0.557$ for scheme (i), whereas we have  
$p_c^{AF1} =0.551$ and $p_c^{AF2}=0.548$ for scheme (ii).  
Again, results of scheme (i) ought to be more accurate than those of
scheme (ii), although the agreement between both schemes is 
a good check of the consistency of our results. We conclude that
the  spin-half $J_1$--$J_2$ HAFM on the SC lattice possesses an 
intermediate quantum phase between two semi-classical magnetic 
phases with continuous transitions between the phases.      
Again, this behaviour is highly reminiscent of the behaviour 
for the order parameter of the spin-half square-lattice $J_1$--$J_2$ model,
as is shown by the inset to Fig. \ref{fig6}. However, the
intermediate quantum paramagnetic regime is much clearer for this 2D model. 
Thus, our data obtained by a high-order CCM approximation provide serious
indications, but not definite evidence, for the presence of the intermediate
quantum phase for the SC lattice.

Results for the spin-spin correlation functions are shown in Fig.~\ref{fig7}, where
CCM results are again in good agreement with results of ED ($N=32$).
The overall shape of the  correlation functions around $p=0.55$ is in a
accordance 
with a continuous transition. Their behavior   
is quite different to the results for the
BCC  model. For example, results for the  nearest-neighbor and next-nearest-neighbor correlation functions 
demonstrate a large discontinuity in values in the region of transition
(centered on $p_c \approx 0.53$) for the BCC lattice, as shown in Fig.~\ref{fig4}. 
By contrast, the changes in the spin-spin correlation functions for the SC lattice  
near the phase transition points are clearly of smaller magnitude and are 
much smoother than for the BCC lattice, as shown in Fig.~\ref{fig7} for both the
ED and CCM results. 

\begin{figure}
\epsfxsize=12cm
\centerline{\epsffile{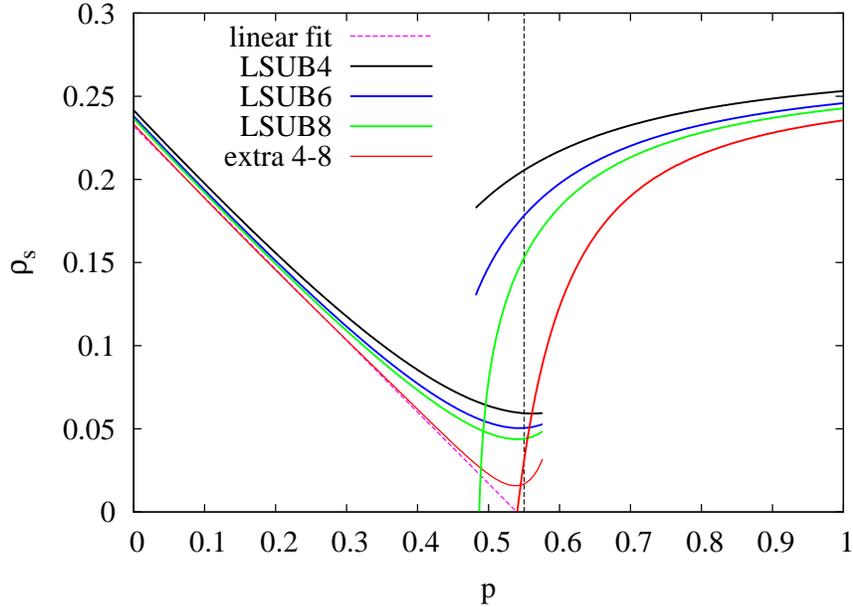}}
\caption{CCM results for the spin stiffness for the spin-half $J_1$--$J_2$ model on the SC lattice.
Extrapolated results (label `extra 4-8') are obtained by using the
extrapolation scheme
of Eq.~(\ref{stiff_extrapo}) using data from the LSUB4, LSUB6, and LSUB8
approximations.
For the classical model we have $\rho^{AF1}_{s,cl} = s^2 \left(J_1 -
4J_2\right)$ and  $\rho^{AF2}_{s,cl} = s^2 J_1$.}
\label{fig8}
\end{figure}

In addition to the sublattice magnetization $M$ we can also use the spin
stiffness $\rho_s$ (see App.~\ref{app_rho}) 
to get an independent
analysis of  order-disorder quantum phase transitions.   
A positive value of $\rho_s$ means that there is magnetic long-range order in the system,
whereas a value of zero reveals that there is no magnetic  long-range order.
Results for $\rho_s$ of the spin-half $J_1$--$J_2$ model  on the SC 
lattice are given in Fig.~\ref{fig8}. 
For the unfrustrated  SC Heisenberg antiferromagnet, i.e. at the point
$p=0$,
we found 
$\rho^{AF1}_s = 0.24158$,   $0.23803$,  $0.23654$
at the LSUB4, LSUB6, and LSUB8 levels of approximation. (Note that the
LSUB4 and  LSUB6 data coincide with those of Ref.~\cite{krueger}, whereas the LSUB8
result is new).
The extrapolated result is  $\rho^{AF1}_s(p=0)=0.2332$, that is close to the result
of Ref.~\cite{krueger} obtained without  LSUB8. 
We also mention that the CCM value  $\rho^{AF1}_s(p=0)=0.2332$ is in very good
agreement with $\rho^{AF1}_s(p=0)=0.2343$ obtained by
second-order spin-wave theory \cite{Hamer1994}. 

At small values of $p$ the stiffness   $\rho^{AF1}_s$ decreases linearly
with increasing $p$. That is similar to the classical result
$\rho^{AF1}_{s,cl}(p)=\rho^{AF1}_{s,cl}(p=0) - bp$, however with a reduced
slope of $b=0.43$ instead of  $b=0.5$.   
Approaching the transition point $p_c$ we find a slight upturn in
$\rho^{AF1}_s$, and, as a result, we cannot determine a transition point by
$\rho^{AF1}_s$. We argue, that  likely higher orders of LSUB$m$ approximations are required to
overcome this problem. However, we may speculate that the linear relation
$\rho^{AF1}_{s}(p)$ (valid at small $p$) remains approximately valid until
$p_c$. A corresponding extrapolation (see the dashed magenta line in
Fig.~\ref{fig8}) crosses the $x$-axis at $p=0.540$, i.e. close the the $p_c$
value found from the order parameter $M$, see Fig.~\ref{fig6}.

In the AF2 phase at larger values of the frustration parameter $p$ the
stiffness $\rho^{AF2}_{s}(p)$ behaves quite differently.
Asymptotically  it saturates as $ p\to \infty$ (note that
$\rho^{AF2}_{s,cl}(p)=const.$). As approaching $p_c$ from the right,
$\rho^{AF2}_{s}(p)$  drops down and the stiffness extrapolated according to
Eq.~(\ref{stiff_extrapo}) vanishes at  $p=0.540$, i.e. at that
value, where the linear fit of $\rho^{AF1}_{s}(p)$ becomes zero.
We remark that $\rho^{AF2}_{s}$ and the linear fit of $\rho^{AF1}_{s}(p)$ both tend to zero at a value of $p$ that is consistent with results for the vanishing points of the order parameter $M$ using model states AF1 and AF2. All of these results demonstrate that the transition is different to that for the BCC lattice. Furthermore, the behavior of the ground-state energy, the order parameter, and the stiffness are quite similar to that found for the square-lattice $J_1$-$J_2$ HAFM \cite{ref013}, albeit with an intermediate quantum phase that is much smaller for the 3D SC lattice.

\section{Conclusions}

The ground-state phases of the spin-half $J_1$--$J_2$ HAFM on the BCC and SC lattices
were
investigated by using the CCM in this article. 
Two antiferromagnetic regimes of collinear order were observed for the BCC lattice, namely, of 
nearest-neighbor N\'eel and next-nearest-neighbor N\'eel striped long-range order. An intersection
point between the ground-state energies for these two model states was observed 
at $p = 0.528$ [where $p = (z_2 J_2) / (z_1 J_1)$], and no intermediate
magnetically disordered phase 
was detected. The gradient of the ground-state energy with respect to $p$ 
(and also for the spin-spin correlation functions using ED) behaved discontinuously at the intersection point. The
values for the corresponding order parameters at this point are 
$M \sim 0.36 \sim 0.41$. These results are all clear indications of a single first-order 
phase transition occurring at $p\sim 0.53$, which is in agreement with results
of all other approximate methods \cite{ref23,ref24,ref25,ref26} applied to this model. 

The spin-half $J_1$--$J_2$ HAFM on the SC lattice is more strongly frustrated
due
to the self-frustrating character of the $J_2$ bonds.
Although the data for the SC lattice were harder to resolve, our results
demonstrate that the ground-state phase diagram is
very different to that of the BCC lattice.
In particular, the investigation of the magnetic order parameter indicates  that there is an 
intermediate quantum phase in between the two semi-classical magnetic
phases.
Thus, the phase diagram of the spin-half $J_1$--$J_2$ HAFM on the SC
lattice resembles that of the corresponding 2D model.
Trivially, any investigation of a highly non-trivial quantum many-body system 
relies on approximations. Bearing in mind that we find  a very small parameter region where this quantum phase may
exist, we cannot exclude that the actual phase diagram does not exhibit such
a quantum phase.
However, our data provide evidence that the quantum  $J_1$--$J_2$ model on the SC
lattice is a candidate for a 3D spin system, where strong frustration may lead 
to a non-magnetic quantum ground state. Moreover, any additional competing 
term in the Hamiltonian would further open the window for an
unconventional quantum phase.  

Evidence in the literature relating to the existence of the intermediate quantum phase for the SC lattice is mixed, and certainly there is no consensus as to its nature, if indeed it does exist. However, there are some similarities between the behavior of the $J_1$-$J_2$ model on the SC lattice and that of the square-lattice $J_1$-$J_2$ model. It is worth noting that calculations for the square-lattice model using density matrix renormalization group with explicit implementation of SU(2) spin rotation symmetry in Ref. \cite{ref020} have found a gapless spin liquid for $0.44 < J_2/J_1 < 0.5$ and a gapped plaquette valence-bond phase for $0.5 < J_2/J_1 < 0.61$. However, any inference relating to the ground-state ordering of the SC-lattice model in the intermediate regime based on the behavior of the square-lattice model would be highly speculative. Bearing in mind that the region of a possible intermediate phase is very small, it seems that the emergence of a sizable gap in this phase is unlikely, i.e., we may expect that the intermediate phase is either a gapless spin liquid or a phase with a very small gap, cf. also the discussion in Ref. \cite{thomale2015}. 

\appendix
\renewcommand{\thefigure}{A\arabic{figure}}
\setcounter{figure}{0}

\section{The spin stiffness of the SC $J_1$-$J_2$ antiferromagnet} 
\label{app_rho}
The spin
stiffness $\rho_s$ measures the increase in the amount of energy as we twist the
magnetic order parameter of a magnetically long-range ordered system along a
given direction by a small angle $\theta$ per unit length, see, e.g.,
Refs.~\cite{schulz1995,lhuillier1995,trumper1998,bishop2015,goetze2016}. We use here the
notations given in Ref.~\cite{trumper1998}
and define the stiffness tensor as 
\begin{equation}
\label{RHO}\rho^{\alpha \beta }=\left.
\frac{\partial ^2e_{%
{g}}({\bf Q)}}{\partial \theta _\alpha \partial \theta _\beta }
\right|_{{\bf Q=}0}, 
\end{equation}
where $e_{{g}}=E_g/N$ is the ground-state energy per spin,
$\theta _\alpha
={\bf Q\cdot e}_\alpha $ ($\alpha $$=1,2,3$) are the twist angles along the
basis vectors ${\bf e}_\alpha $, and
$\bf Q$ is the magnetic wave vector of the magnetically long-range ordered
phase.

For the SC lattice we have trivially ${\bf e}_\alpha={\bf e}_{x,y,z}$. The
coresponding magnetic wave-vectors are ${\bf Q} =(\pi,\pi,\pi)$ for the
AF1 (N\'eel) state (see Fig.~\ref{fig1}a) and  ${\bf Q} =(\pi,0,\pi)$
for the AF2 (striped) state (see Fig.~\ref{fig1}b).
For the classical model in the AF1 phase we easily obtain
 $\rho_{cl}^{\alpha \beta } = 
\rho^{AF1}_{s,cl}\delta_{\alpha \beta}$ with $\rho^{AF1}_{s,cl} = s^2 \left(J_1 - 4J_2\right)$, 
i.e. the stiffness tensor is diagonal and naturally the $x$, $y$
and $z$-components are identical.

For the magnetic wave vector ${\bf
Q}=(\pi,0,\pi)$ (AF2 state) 
we have  to consider the 
twists $\theta_\alpha
={\bf Q\cdot e}_\alpha $, i.e.  $\theta_x=\theta_1$,  $\theta_y=0$, ,
$\theta_z=\theta_2$, and we obtain for the classical
model again a diagonal tensor 
 $\rho_{cl}^{\alpha \beta } = 
\rho^{AF2}_{s,cl}\delta_{\alpha \beta}$ with $\rho^{AF2}_{s,cl} = s^2 J_1$.
The CCM calculation for the quantum $s=1/2$ model is straightforward, see
Refs.~\cite{ref013,krueger,goetze2016,bishop2015}. 
We introduce the twist as described above and use the twisted state as the
model state for the CCM calculation. As a result we obtain the quantum
ground-state energy as a function of the imposed twist angle that can be used to
find $\rho^{AF1}_s$ and $\rho^{AF2}_s$ according to Eq.~(\ref{RHO}).   
However, note that the solution of the corresponding CCM-LSUB$m$ equations
is more challenging because fewer point-group symmetries can be used for the 
non-collinear twisted state and so we have more fundamental clusters at 
equivalent level of LSUB$m$ approximation. Therefore we can only calculate 
the stiffness only up  to LSUB8.   
We follow Refs.~\cite{ref013,krueger,goetze2016,bishop2015} and extrapolate the stiffness CCM-LSUB$m$
data to $m \to \infty$ using LSUB4, LSUB6 and LSUB8 data by using the extrapolation scheme
given by 
\begin{equation}
\rho_s(m) = \rho_s(m = \infty) + c_{1}m^{-1}+c_{2}m^{-2}\, .
\label{stiff_extrapo}
\end{equation}

\end{document}